\documentstyle[12pt,epsfig]{article}
\begin{document}
\begin{flushright}
Fermilab-PUB-99/360-E\\
CDF/PUB/JET/PUBLIC/4760 \\
December 10, 1999\\
\end{flushright}
\vspace{0.5in}
\begin{center}
\begin{Large}
{\bf  A Measurement of the Differential Dijet Mass Cross Section in 
$p\bar{p}$ Collisions at $\sqrt{s}=1.8$ TeV}
\end{Large}
\end{center}
\vspace{0.5in}
\font\eightit=cmti8
\def\r#1{\ignorespaces $^{#1}$}
\hfilneg
\begin{sloppypar}
\noindent
T.~Affolder,\r {21} H.~Akimoto,\r {43}
A.~Akopian,\r {36} M.~G.~Albrow,\r {10} P.~Amaral,\r 7 S.~R.~Amendolia,\r {32} 
D.~Amidei,\r {24} K.~Anikeev,\r {22} J.~Antos,\r 1 
G.~Apollinari,\r {36}
% ...
\end{sloppypar}
\vskip .026in
\begin{center}
(CDF Collaboration)
\end{center}

\vskip .026in
\begin{center}
\r 1  {\eightit Institute of Physics, Academia Sinica, Taipei, Taiwan 11529, 
Republic of China} \\
\r 2  {\eightit Argonne National Laboratory, Argonne, Illinois 60439} \\
\r 3  {\eightit Istituto Nazionale di Fisica Nucleare, University of Bologna,
I-40127 Bologna, Italy} \\
\r 4  {\eightit Brandeis University, Waltham, Massachusetts 02254} \\
\r 5  {\eightit University of California at Los Angeles, Los 
Angeles, California  90024} \\  
\r 6  {\eightit Instituto de Fisica de Cantabria, University of Cantabria, 
39005 Santander, Spain} \\
\r 7  {\eightit Enrico Fermi Institute, University of Chicago, Chicago, 
Illinois 60637} \\
\r 8  {\eightit Joint Institute for Nuclear Research, RU-141980 Dubna, Russia}
\\
\r 9  {\eightit Duke University, Durham, North Carolina  27708} \\
\r {10}  {\eightit Fermi National Accelerator Laboratory, Batavia, Illinois 
60510} \\
\r {11} {\eightit University of Florida, Gainesville, Florida  32611} \\
\r {12} {\eightit Laboratori Nazionali di Frascati, Istituto Nazionale di Fisica
               Nucleare, I-00044 Frascati, Italy} \\
\r {13} {\eightit University of Geneva, CH-1211 Geneva 4, Switzerland} \\
\r {14} {\eightit Harvard University, Cambridge, Massachusetts 02138} \\
\r {15} {\eightit Hiroshima University, Higashi-Hiroshima 724, Japan} \\
\r {16} {\eightit University of Illinois, Urbana, Illinois 61801} \\
\r {17} {\eightit The Johns Hopkins University, Baltimore, Maryland 21218} \\
\r {18} {\eightit Institut f\"{u}r Experimentelle Kernphysik, 
Universit\"{a}t Karlsruhe, 76128 Karlsruhe, Germany} \\
\r {19} {\eightit Korean Hadron Collider Laboratory: Kyungpook National
University, Taegu 702-701; Seoul National University, Seoul 151-742; and
SungKyunKwan University, Suwon 440-746; Korea} \\
\r {20} {\eightit High Energy Accelerator Research Organization (KEK), Tsukuba, 
Ibaraki 305, Japan} \\
\r {21} {\eightit Ernest Orlando Lawrence Berkeley National Laboratory, 
Berkeley, California 94720} \\
\r {22} {\eightit Massachusetts Institute of Technology, Cambridge,
Massachusetts  02139} \\   
\r {23} {\eightit Institute of Particle Physics: McGill University, Montreal 
H3A 2T8; and University of Toronto, Toronto M5S 1A7; Canada} \\
\r {24} {\eightit University of Michigan, Ann Arbor, Michigan 48109} \\
\r {25} {\eightit Michigan State University, East Lansing, Michigan  48824} \\
\r {26} {\eightit University of New Mexico, Albuquerque, New Mexico 87131} \\
\r {27} {\eightit The Ohio State University, Columbus, Ohio  43210} \\
\r {28} {\eightit Osaka City University, Osaka 588, Japan} \\
\r {29} {\eightit University of Oxford, Oxford OX1 3RH, United Kingdom} \\
\r {30} {\eightit Universita di Padova, Istituto Nazionale di Fisica 
          Nucleare, Sezione di Padova, I-35131 Padova, Italy} \\
\r {31} {\eightit University of Pennsylvania, Philadelphia, 
        Pennsylvania 19104} \\   
\r {32} {\eightit Istituto Nazionale di Fisica Nucleare, University and Scuola
               Normale Superiore of Pisa, I-56100 Pisa, Italy} \\
\r {33} {\eightit University of Pittsburgh, Pittsburgh, Pennsylvania 15260} \\
\r {34} {\eightit Purdue University, West Lafayette, Indiana 47907} \\
\r {35} {\eightit University of Rochester, Rochester, New York 14627} \\
\r {36} {\eightit Rockefeller University, New York, New York 10021} \\
\r {37} {\eightit Rutgers University, Piscataway, New Jersey 08855} \\
\r {38} {\eightit Texas A\&M University, College Station, Texas 77843} \\
\r {39} {\eightit Texas Tech University, Lubbock, Texas 79409} \\
\r {40} {\eightit Istituto Nazionale di Fisica Nucleare, University of Trieste/
Udine, Italy} \\
\r {41} {\eightit University of Tsukuba, Tsukuba, Ibaraki 305, Japan} \\
\r {42} {\eightit Tufts University, Medford, Massachusetts 02155} \\
\r {43} {\eightit Waseda University, Tokyo 169, Japan} \\
\r {44} {\eightit University of Wisconsin, Madison, Wisconsin 53706} \\
\r {45} {\eightit Yale University, New Haven, Connecticut 06520} \\
\end{center}

\renewcommand{\baselinestretch}{2}
\large
\normalsize

\begin{center}
{\bf Abstract}
\end{center}
We present a measurement of the cross section for production of two or more 
jets as a function of dijet mass, based on an integrated luminosity of 
86 ${\rm pb}^{-1}$ collected with the Collider Detector at Fermilab.
Our dijet mass spectrum is described within errors by next-to-leading order QCD 
predictions using CTEQ4HJ parton distributions, and is in good agreement 
with a similar measurement from the D$\emptyset$ experiment.

\vspace*{0.5in}

\clearpage

Hard collisions between protons and antiprotons predominantly produce 
dijet events, which are events containing at least two high energy jets.
A measurement of the 
dijet mass differential cross section provides a fundamental test of
Quantum Chromodynamics (QCD) 
and a constraint on the parton distributions of the proton.
We previously reported measurements of the inclusive jet transverse energy 
($E_T$) spectrum~\cite{ref_incl_jet} and the cross section for events with 
large total $E_T$~\cite{ref_sum_et}.
Both measurements indicated an excess of events at high $E_T$ compared to the 
predictions of QCD. This letter 
presents our most recent measurement of the dijet mass 
spectrum~\cite{ref_bjoern} and compares 
it with the predictions of next-to-leading order QCD and the measurement of 
D$\emptyset$~\cite{ref_mass_D0}. This measurement, with an integrated 
luminosity of $86$ pb$^{-1}$, is significantly more sensitive to events at high 
dijet mass than our previous measurements of the dijet mass 
spectrum~\cite{ref_old_cdf} with integrated luminosities of $4.2$ 
pb$^{-1}$ and $26$ nb$^{-1}$. 
We recently used this data sample combined with 20 pb$^{-1}$ of older data to 
measure dijet angular distributions~\cite{ref_angle} and to search the dijet 
mass spectrum for new particles decaying to dijets~\cite{ref_search}.

A detailed description of the Collider Detector at Fermilab (CDF) can be found 
elsewhere~\cite{ref_CDF}. We use a coordinate system with the $z$ axis along 
the proton beam, transverse coordinate perpendicular to the beam, azimuthal 
angle $\phi$, polar angle $\theta$, and pseudorapidity 
$\eta=-\ln \tan(\theta/2)$. 
Jets are reconstructed as localized energy depositions in the CDF calorimeters, 
which are arranged in a projective tower geometry.
The jet energy, $E$, is defined as the scalar sum of the calorimeter tower 
energies inside a cone of radius 
$R=\sqrt{(\Delta\eta)^2 + (\Delta\phi)^2}=0.7$, centered on the jet direction. 
Jets that share towers are combined if the total $E_T$ of the 
shared towers is greater than 75\% of the $E_T$ of either jet; otherwise 
the towers are assigned to the nearest jet.
The jet momentum, $\vec{P}$, is the vector sum: 
$\vec{P} = \sum{E_i\hat{u}_i}$, with $\hat{u}_i$ being the 
unit vector pointing from the interaction point to the energy deposition $E_i$ 
inside the cone. The quantities $E$ and $\vec{P}$ are corrected 
for calorimeter non-linearities, energy lost in uninstrumented regions of the
detector, and energy gained from the underlying event and additional $p\bar{p}$
interactions. We do not correct for energy lost outside the clustering cone, 
since a similar loss is present in the O($\alpha_s^3$) QCD calculation in which
an extra gluon can be radiated outside the jet clustering cone.
The jet energy corrections increase the measured jet energies on average by 
20\% (16\%) for 100 GeV (400 GeV) jets.  
Full details of jet reconstruction and
jet energy corrections at CDF can be found elsewhere~\cite{ref_jet}.

We define the dijet system as the
two jets with the highest transverse momentum in an event (leading jets) 
and define the dijet mass as 
$M=\sqrt{(E_1 + E_2)^2 - (\vec{P}_1 + \vec{P}_2)^2}$.  
Our data sample was obtained using four triggers 
that required at least one jet with uncorrected cluster transverse energies
of 20, 50, 70 and 100 GeV, respectively.  After correcting the jet energies these 
trigger samples
were used to measure the dijet mass spectrum above 180, 217, 292, and 388 
GeV/c$^2$, respectively, where the trigger 
efficiencies were greater than 97\%. The four data samples 
corresponded to integrated luminosities of $0.091$, $2.2$, $11$, and $86$ 
pb$^{-1}$ respectively.  We selected events with two or more jets and 
required that the two leading jets have pseudorapidities of $|\eta_1|<2$ and 
$|\eta_2|<2$ and satisfy 
$|\cos\theta^*| = |\tanh[(\eta_1-\eta_2)/2]| < 2/3$, where $\theta^*$ 
is the scattering angle in the dijet center-of-mass frame.
The $\cos\theta^*$ 
requirement ensures full acceptance as a function of the dijet mass.
The $z$ position of the event vertex was required to be within 60 cm of the 
center of the detector; this cut removed 6\% of the events. 
Backgrounds from cosmic rays, 
beam halo, and detector noise were removed 
by requiring \mbox{${\not\!\!E_T}/\sqrt{\sum E_T}<6$} GeV$^{1/2}$ and 
$\sum E< 2$ TeV,
where \mbox{${\not\!\!E_T}$} is the missing transverse
energy~\cite{ref_miss_et}, $\sum E_T$ is the total transverse energy (scalar 
sum), and $\sum E$ is the total
energy in the event. These cuts selected 60,998 events.

The dijet mass resolution was determined using the 
PYTHIA~\cite{ref_PYTHIA} Monte Carlo program and a CDF detector simulation.
The true jet is defined from the true $E_T$ of particles emanating from the
hard scattering, using the same jet algorithm as described above, but applied
to towers of true $E_T$. The true $E_T$ of a tower is the $E_T$ of the 
generated particles that enter the tower.
The simulated jet uses the $E_T$ of simulated 
calorimeter towers and the jet energy corrections for the CDF detector 
simulation. The $E_T$ of the simulated jets is corrected to equal the 
$E_T$ of the corresponding true jet on average. The dijet mass resolution
function, $\rho(M,m)$, is then defined as the distribution of simulated dijet 
masses, $M$,
for each value of true dijet mass, $m$. The dijet mass resolution was determined 
for six values of $m$ between 50 and 1000 GeV/c$^2$ and then 
a single smooth parameterization was used to interpolate between these values.
The dijet mass resolution is approximately 10\% for dijet masses 
above 150 GeV/c$^2$.

The steeply falling dijet mass spectrum is distorted by the dijet mass 
resolution. 
We correct for this distortion with an unsmearing procedure.
Define the smeared spectrum, $S(M)$, as the convolution of the true spectrum, 
$T(m)$, and the dijet mass resolution: $S(M) = \int T(m) \rho(M,m) dm$. 
We parameterize the true dijet mass 
spectrum with $T(m) = A(1 - m/\sqrt{s} + Cm^2/s)^N/m^P$ 
where $\sqrt{s} = 1800$ GeV.
We then fit the smeared spectrum to our data to find the value of the 
four parameters $A=6.67\times 10^{17}$pb/(GeV/c$^2$), $C=2.95$, $N=-6.98$, 
and $P=6.70$. The fit has a $\chi^2$ of 20.5 for 14 degrees of freedom.
The unsmearing correction factors, $K_i$, are then defined as the 
ratio of the smeared to true spectrum, 
$K_i = \int_i S(M) dM / \int_i T(m)dm$, where 
the integration is over mass bin $i$.  The value of $K_i$
smoothly decreases from $1.07$ at $M=188$ GeV/c$^2$, to $1.03$ at 
$M=540$ GeV/c$^2$, and then smoothly increases to $1.12$ at $M=968$ GeV/c$^2$.
The corrected cross section as a function of dijet mass is given by 
\begin{equation}
d\sigma/dM = n_i /(K_i \ {\mathcal L} \ \epsilon_i \ \Delta M),
\end{equation}
where for each mass bin $i$, $n_i$ is the number of events, $\mathcal{L}$ is the 
integrated luminosity, $\epsilon_i$ is the efficiency of the 
trigger and $z$-vertex selections, and $\Delta M$ is the width of the mass bin.

In Table I we list 12 independent sources of systematic uncertainty in the
dijet mass cross section. They are the
uncertainties in calorimeter calibration (cal), jet fragmentation (frag), 
underlying event (uevt), calorimeter stability over time (stab), relative jet 
energy scale as a function of pseudorapidity~\cite{ref_angle} (rel), 
detector simulation (sim), the unsmearing  procedure (unsm), 
the tails of the resolution function (tails), 
absolute 
luminosity~\cite{ref_lum} of the jet 100 trigger (lum), and the 
relative luminosities of the jet 20, 50 and 70 triggers (J20, J50, and J70).
The first four systematic uncertainties~\cite{ref_incl_jet} are 
equivalent to a combined uncertainty in the determination of the dijet mass 
variable which decreases from 2.7\% at $M=188$ GeV/c$^2$ to 2.3\% at 
$M=968$ GeV/c$^2$. 
The uncertainty in detector simulation results from a $0.5$\% uncertainty 
in the equality of the true dijet mass and the simulated dijet mass after all jet 
corrections are applied, independent from the first four systematic 
uncertainties mentioned above.
To check that our unsmearing procedure is internally consistent, 
we applied the unsmearing procedure to a simulated dijet mass spectrum.
The resulting $K_i$ were in agreement with the ratios of the simulated spectrum
to true spectrum for each mass bin. Due to limited Monte Carlo statistics, the
systematic uncertainty on the consistency of the unsmearing procedure was 4\%.
The uncertainty in the dijet mass resolution due to non-Gaussian tails was
estimated by repeating the unsmearing procedure with a Gaussian resolution.
The systematic uncertainties on the luminosity for the jet 20, 50 and 70 
triggers came from the statistical uncertainty in matching the cross section
of each trigger with the next higher threshold trigger (jet 70 was required
to match jet 100 in the first bin of the jet 100 sample, jet 50 was required 
to match jet 70, etc.)
Each of the independent systematic uncertainties in Table I are completely 
correlated as a function of dijet mass.  

\begin{figure}[tbh]
\centerline{\epsfig{figure=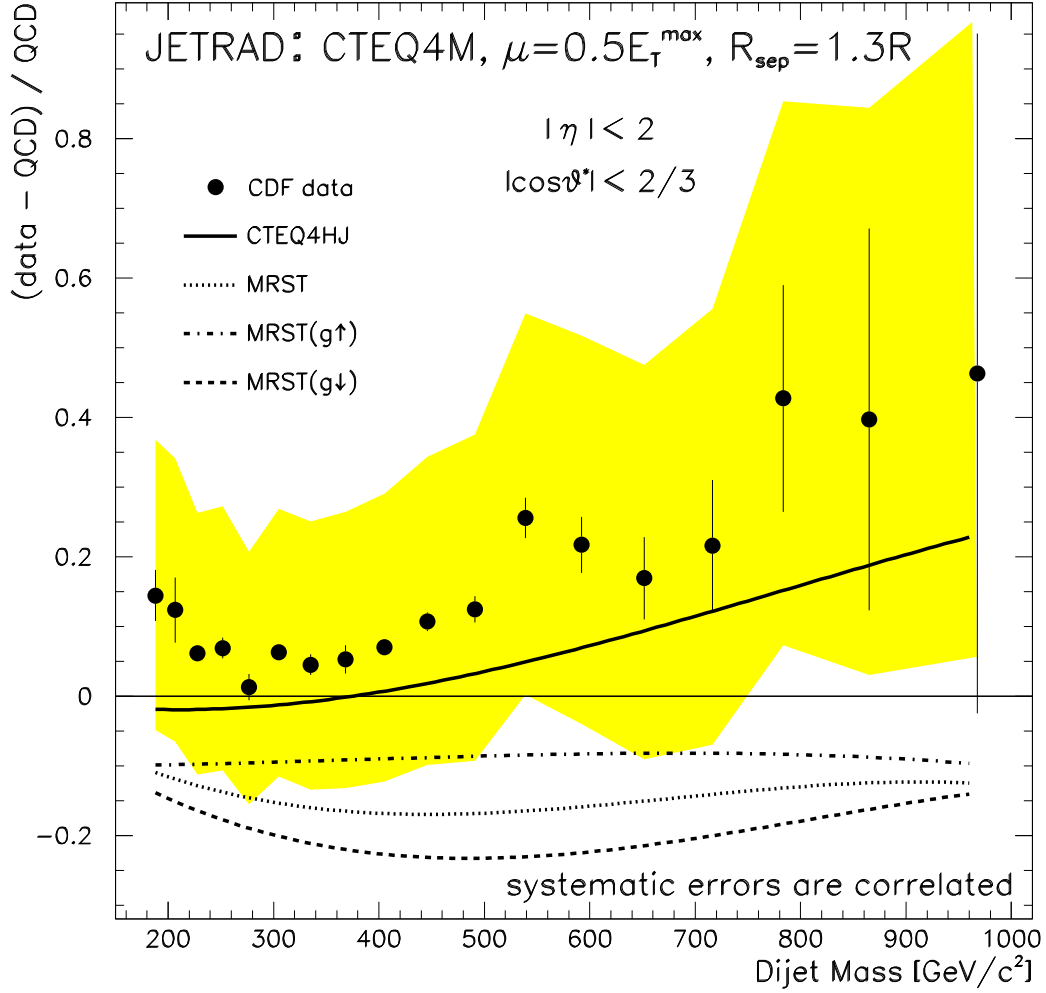,width=6in}}
\caption{ 
The fractional difference between the measured differential cross section and 
the QCD prediction (points) as a function of dijet mass. The band is the
systematic uncertainty. The curves are the fractional difference between other 
QCD predictions, for various choices of parton distributions, 
and our default QCD prediction using CTEQ4M.}
\label{fig_lin}
\end{figure}

\begin{figure}[tbh]
\centerline{\epsfig{figure=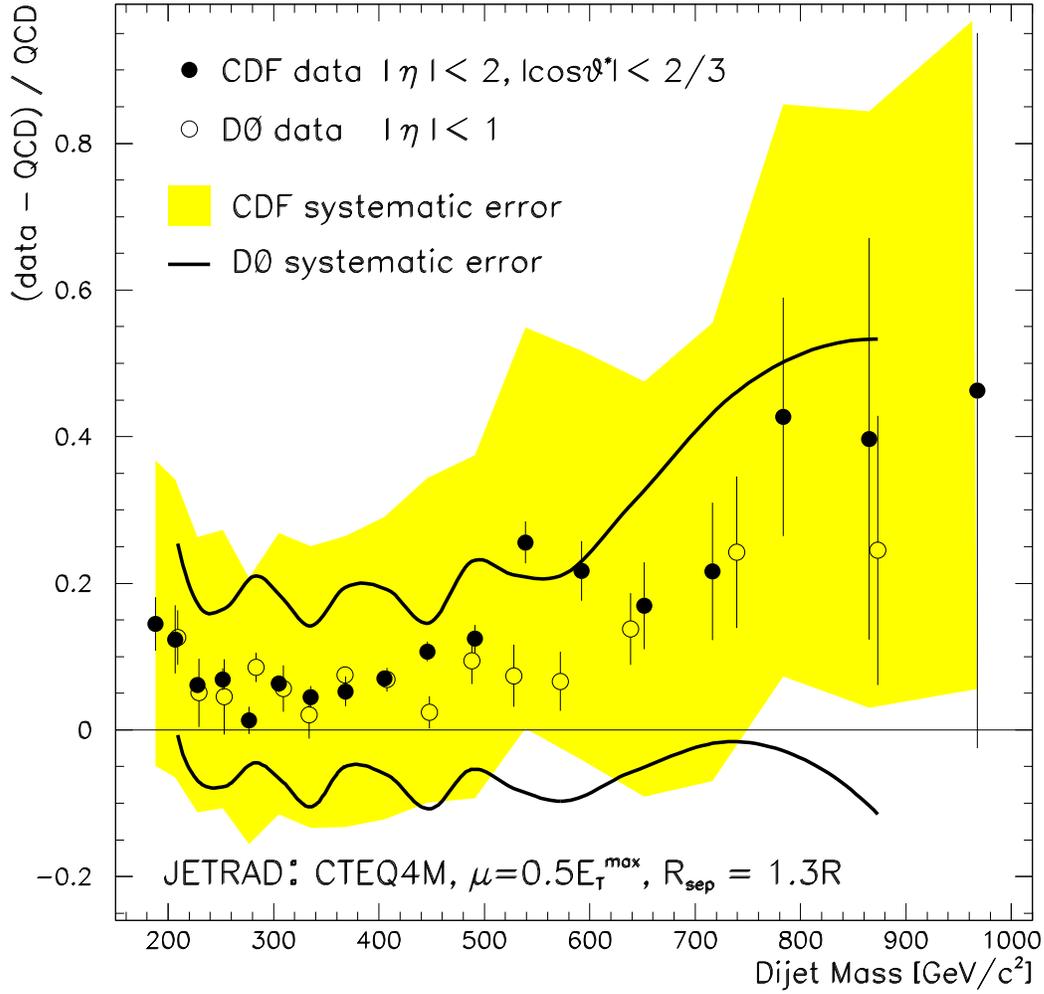,width=6in}}
\caption{ 
The difference between CDF data and QCD (solid points as shown in 
Fig.~\ref{fig_lin}) compared to the difference between 
D$\emptyset$ data~\cite{ref_mass_D0} and QCD (open points).  The solid curves are the D$\emptyset$ systematic 
uncertainty.}
\label{fig_d0}
\end{figure}

In Table II we present the fully corrected 
inclusive dijet 
mass spectrum for $p\bar{p}\rightarrow$ 2 jets + X, where X can be anything, 
including additional jets. 
We tabulate the differential cross section versus the mean dijet mass in 
bins of width approximately equal to the dijet mass resolution. 
Figure~\ref{fig_lin} shows the fractional difference between our data and
O($\alpha_s^3$) QCD predictions from the parton level event 
generator JETRAD~\cite{ref_jetrad}. Here the renormalization scale is 
$\mu=0.5E_T^{max}$, where $E_T^{max}$ is the maximum jet $E_T$ in the generated 
event.  In the JETRAD calculation, two partons are combined if they are 
within $R_{sep} = 1.3 R$, which corresponds to the separation of jets in the 
data.
Predictions are shown for various choices of parton distribution functions: 
CTEQ4M~\cite{ref_CTEQ} and MRST~\cite{ref_MRST} are standard sets and
CTEQ4HJ~\cite{ref_CTEQ} adjusts the gluon distribution to give a better fit 
to the CDF inclusive jet $E_T$ spectrum at high $E_T$. 
Figure~\ref{fig_lin} shows that the CTEQ4HJ 
prediction models the shape and normalization of our dijet data better than 
CTEQ4M. The CTEQ4M prediction changes by less than 5\% when the renormalization
scale is changed to $\mu=E_T^{max}$, but it decreases between 7\% and 17\% for 
$\mu=2E_T^{max}$, and it decreases between 25\% and 30\% for $\mu=0.25E_T^{max}$.
In Fig.~\ref{fig_d0} we compare 
the fractional difference between our data and QCD with that of the D$\emptyset$ 
experiment. The D$\emptyset$ measurement~\cite{ref_mass_D0} and the JETRAD prediction 
obtained by D$\emptyset$ required that each jet be in region 
$|\eta|<1.0$.
Figure~\ref{fig_d0} shows that our data and the D$\emptyset$ data are in good agreement. 

The covariance matrix for the dijet mass differential cross section is
defined as $V_{ij} = \delta_{ij}\sigma_i^2(stat) + \Sigma_{k=1}^{12} \sigma_i(sys_k) 
\sigma_j(sys_k)$.
Here $\delta_{ij}=1(0)$ for $i=j(i\neq j)$, $\sigma_i(stat)$ is the 
statistical uncertainty in mass bin $i$, and the sum is over each of the 12  
systematic uncertainties $\sigma_i(sys_k)$ listed in Table I. Since the theory
always predicts a smaller cross section than the data, the positive 
percent systematic uncertainty given in Table I was multiplied by the 
theoretical cross section to determine the $\sigma_i(sys_k)$.
From the inverse of the covariance 
matrix, $(V^{-1})_{ij}$, and the difference between the data and the theory in 
each bin, $\Delta_i$,  we perform a $\chi^2$ comparison between the data
and the theory. Table III presents values for 
$\chi^2 = \Sigma_{i,j} \Delta_i (V^{-1})_{ij} \Delta_j$ and the corresponding
probability for a standard $\chi^2$ distribution with 18 degrees of freedom
(14 degrees of freedom for the row labeled Fit). Our data is in 
agreement within errors with the QCD prediction 
using CTEQ4HJ parton distributions, which has an enhanced gluon distribution 
at high $E_T$. Our data excludes CTEQ4M parton distributions, 
which have a standard gluon distribution.  
The $\chi^2$ comparison shows that our data cannot exclude with high 
confidence QCD predictions 
using MRST parton distributions, even though the normalization of that 
prediction is well beneath that of our data. 
This is because of the presence of correlated systematic 
uncertainties that are large compared with the statistical 
uncertainties. Such correlated uncertainties can accommodate certain 
significant deviations in both normalization and shape between the data 
and the theory with a relatively small penalty in $\chi^2$. Any 
theoretical prediction whose deviation from the data matches the shape 
of a correlated uncertainty will give a reasonable $\chi^2$ provided that the 
normalization difference between the data and the prediction is no more 
than a few standard deviations.

In conclusion, we have measured the cross section for production of two 
or more jets in the kinematic region $|\eta|<2$ and $|\cos\theta^*|<2/3$ 
as a function of dijet invariant mass. The data at the highest values of dijet 
mass are above the QCD predictions using standard parton distributions, similar 
to the excess at high $E_T$ observed in previous measurements of the inclusive 
jet $E_T$ spectrum~\cite{ref_incl_jet} and the total 
$E_T$ spectrum~\cite{ref_sum_et}. 
The CDF data are described within errors by next-to-leading order QCD 
predictions using CTEQ4HJ parton distributions, and are in good agreement 
with a similar measurement from the D$\emptyset$ experiment.

We thank the Fermilab staff and the technical staffs of the participating 
institutions for their vital contributions. This work was
supported by the U.S. Department of Energy and National Science Foundation;
the Italian Istituto Nazionale di Fisica Nucleare; the Ministry of Education, 
Science and Culture of Japan; the Natural Sciences and Engineering Research
Council of Canada; the National Science Council of the Republic of China; 
the Swiss National Science Foundation; the A. P. Sloan Foundation; the
Bundesministerium fuer Bildung und Forschung, Germany; and the Korea Science 
and Engineering Foundation.

\clearpage

\renewcommand{\baselinestretch}{1.4}
\large
\normalsize

\begin{table}[tbh]
Table I: Systematic uncertainties on the cross section (see text). 
\renewcommand{\baselinestretch}{1.2}
\begin{center}
\begin{tabular}{|c|c|c|c|c|c|c|c|c|c|c|c|c|}\hline
Mass & \multicolumn{12}{c|}{Systematic Uncertainty on Cross Section in \%}\\ \cline{2-13}
(GeV/c$^2$) & cal & frag & uevt & stab & rel & sim & unsm & tails & lum & J20 & J50 &
J70 \\ \hline

188 &
{\footnotesize $\begin{array}{l}\raisebox{-.05cm}{+12}\\ 
\raisebox{0.1cm}{-- \ 8}\end{array}$}
&
{\footnotesize $\begin{array}{l}\raisebox{-.05cm}{+ 8}\\ 
\raisebox{0.1cm}{-- \ 7}\end{array}$}
&
{\footnotesize $\begin{array}{l}\raisebox{-.05cm}{+ 8}\\ 
\raisebox{0.1cm}{-- \ 7}\end{array}$}
&
{\footnotesize $\begin{array}{l}\raisebox{-.05cm}{+ 7}\\ 
\raisebox{0.1cm}{-- \ 6}\end{array}$}
& 5 & 2 & 4 & 2 & 4 & 4 & 2 & 2
\\ \hline

207 &
{\footnotesize $\begin{array}{l}\raisebox{-.05cm}{+12}\\ 
\raisebox{0.1cm}{-- \ 8}\end{array}$}
&
{\footnotesize $\begin{array}{l}\raisebox{-.05cm}{+ 8}\\ 
\raisebox{0.1cm}{-- \ 7}\end{array}$}
& 
7 
&
{\footnotesize $\begin{array}{l}\raisebox{-.05cm}{+ 7}\\ 
\raisebox{0.1cm}{-- \ 6}\end{array}$}
& 5 & 2 & 4 & 2 & 4 & 4 & 2 & 2
\\ \hline

228 &
{\footnotesize $\begin{array}{l}\raisebox{-.05cm}{+12}\\ 
\raisebox{0.1cm}{-- \ 8}\end{array}$}
&
{\footnotesize $\begin{array}{l}\raisebox{-.05cm}{+ 8}\\ 
\raisebox{0.1cm}{-- \ 7}\end{array}$}
&
6
&
{\footnotesize $\begin{array}{l}\raisebox{-.05cm}{+ 7}\\ 
\raisebox{0.1cm}{-- \ 6}\end{array}$}
&
{\footnotesize $\begin{array}{l}\raisebox{-.05cm}{+ 6}\\ 
\raisebox{0.1cm}{-- \ 5}\end{array}$}
& 3 & 4 & 3 & 4 & -- & 2 & 2
\\ \hline

252 &
{\footnotesize $\begin{array}{l}\raisebox{-.05cm}{+12}\\ 
\raisebox{0.1cm}{-- \ 8}\end{array}$}
&
{\footnotesize $\begin{array}{l}\raisebox{-.05cm}{+ 8}\\ 
\raisebox{0.1cm}{-- \ 7}\end{array}$}
&
6
&
{\footnotesize $\begin{array}{l}\raisebox{-.05cm}{+ 7}\\ 
\raisebox{0.1cm}{-- \ 6}\end{array}$}
& 6 & 3 & 4 & 3 & 4 & -- & 2 & 2
\\ \hline

277 &
{\footnotesize $\begin{array}{l}\raisebox{-.05cm}{+12}\\ 
\raisebox{0.1cm}{-- \ 8}\end{array}$}
&
{\footnotesize $\begin{array}{l}\raisebox{-.05cm}{+ 8}\\ 
\raisebox{0.1cm}{-- \ 7}\end{array}$}
&
5
&
{\footnotesize $\begin{array}{l}\raisebox{-.05cm}{+ 7}\\ 
\raisebox{0.1cm}{-- \ 6}\end{array}$}
& 6 & 3 & 4 & 4 & 4 & -- & 2 & 2
\\ \hline

305 &
{\footnotesize $\begin{array}{l}\raisebox{-.05cm}{+12}\\ 
\raisebox{0.1cm}{-- \ 8}\end{array}$}
&
8
&
5
&
7
& 6 & 4 & 4 & 4 & 4 & -- & -- & 2
\\ \hline

335 &
{\footnotesize $\begin{array}{l}\raisebox{-.05cm}{+12}\\ 
\raisebox{0.1cm}{-- \ 8}\end{array}$}
&
{\footnotesize $\begin{array}{l}\raisebox{-.05cm}{+ 8}\\ 
\raisebox{0.1cm}{-- \ 7}\end{array}$}
&
{\footnotesize $\begin{array}{l}\raisebox{-.05cm}{+ 5}\\ 
\raisebox{0.1cm}{-- \ 4}\end{array}$}
&
{\footnotesize $\begin{array}{l}\raisebox{-.05cm}{+ 7}\\ 
\raisebox{0.1cm}{-- \ 6}\end{array}$}
& 6 & 4 & 4 & 5 & 4 & -- & -- & 2
\\ \hline

368 &
{\footnotesize $\begin{array}{l}\raisebox{-.05cm}{+13}\\ 
\raisebox{0.1cm}{-- \ 8}\end{array}$}
&
{\footnotesize $\begin{array}{l}\raisebox{-.05cm}{+ 8}\\ 
\raisebox{0.1cm}{-- \ 8}\end{array}$}
&
{\footnotesize $\begin{array}{l}\raisebox{-.05cm}{+ 5}\\ 
\raisebox{0.1cm}{-- \ 4}\end{array}$}
&
7
&
{\footnotesize $\begin{array}{l}\raisebox{-.05cm}{+ 7}\\ 
\raisebox{0.1cm}{-- \ 6}\end{array}$}
& 5 & 4 & 5 & 4 & -- & -- & 2
\\ \hline
405 &
{\footnotesize $\begin{array}{l}\raisebox{-.05cm}{+13}\\ 
\raisebox{0.1cm}{-- \ 9}\end{array}$}
&
8
&
4
&
{\footnotesize $\begin{array}{l}\raisebox{-.05cm}{+ 8}\\ 
\raisebox{0.1cm}{-- \ 7}\end{array}$}
&
{\footnotesize $\begin{array}{l}\raisebox{-.05cm}{+ 7}\\ 
\raisebox{0.1cm}{-- \ 6}\end{array}$}
& 5 & 4 & 5 & 4 & -- & -- & --
\\ \hline

446 &
{\footnotesize $\begin{array}{l}\raisebox{-.05cm}{+14}\\ 
\raisebox{0.1cm}{-- \ 9}\end{array}$}
&
{\footnotesize $\begin{array}{l}\raisebox{-.05cm}{+ 9}\\ 
\raisebox{0.1cm}{-- \ 8}\end{array}$}
&
4
&
{\footnotesize $\begin{array}{l}\raisebox{-.05cm}{+ 8}\\ 
\raisebox{0.1cm}{-- \ 7}\end{array}$}
&
7 
& 5 & 4 & 6 & 4 & -- & -- & --
\\ \hline

491 &
{\footnotesize $\begin{array}{l}\raisebox{-.05cm}{+14}\\ 
\raisebox{0.1cm}{-- \ 9}\end{array}$}
&
{\footnotesize $\begin{array}{l}\raisebox{-.05cm}{+ 9}\\ 
\raisebox{0.1cm}{-- \ 8}\end{array}$}
&
4
&
8
&
7
& 6 & 4 & 6 & 4 & -- & -- & --
\\ \hline

539 &
{\footnotesize $\begin{array}{l}\raisebox{-.05cm}{+15}\\ 
\raisebox{0.1cm}{--\ 10}\end{array}$}
&
{\footnotesize $\begin{array}{l}\raisebox{-.05cm}{+10}\\ 
\raisebox{0.1cm}{-- \ 9}\end{array}$}
&
{\footnotesize $\begin{array}{l}\raisebox{-.05cm}{+ 4}\\ 
\raisebox{0.1cm}{-- \ 3}\end{array}$}
&
{\footnotesize $\begin{array}{l}\raisebox{-.05cm}{+ 9}\\ 
\raisebox{0.1cm}{-- \ 8}\end{array}$}
&
{\footnotesize $\begin{array}{l}\raisebox{-.05cm}{+ 8}\\ 
\raisebox{0.1cm}{-- \ 7}\end{array}$}
& 6 & 4 & 7 & 4 & -- & -- & --
\\ \hline

592 &
{\footnotesize $\begin{array}{l}\raisebox{-.05cm}{+16}\\ 
\raisebox{0.1cm}{--\ 11}\end{array}$}
&
{\footnotesize $\begin{array}{l}\raisebox{-.05cm}{+10}\\ 
\raisebox{0.1cm}{-- \ 9}\end{array}$}
&
3
&
9
&
{\footnotesize $\begin{array}{l}\raisebox{-.05cm}{+ 8}\\ 
\raisebox{0.1cm}{-- \ 7}\end{array}$}
& 6 & 4 & 7 & 4 & -- & -- & --
\\ \hline

652 &
{\footnotesize $\begin{array}{l}\raisebox{-.05cm}{+17}\\ 
\raisebox{0.1cm}{--\ 11}\end{array}$}
&
{\footnotesize $\begin{array}{l}\raisebox{-.05cm}{+11}\\ 
\raisebox{0.1cm}{--\ 10}\end{array}$}
&
3
&
{\footnotesize $\begin{array}{l}\raisebox{-.05cm}{+10}\\ 
\raisebox{0.1cm}{-- \ 9}\end{array}$}
&
{\footnotesize $\begin{array}{l}\raisebox{-.05cm}{+ 8}\\ 
\raisebox{0.1cm}{-- \ 7}\end{array}$}
& 7 & 4 & 7 & 4 & -- & -- & --
\\ \hline

716 &
{\footnotesize $\begin{array}{l}\raisebox{-.05cm}{+19}\\ 
\raisebox{0.1cm}{--\ 12}\end{array}$}
&
{\footnotesize $\begin{array}{l}\raisebox{-.05cm}{+12}\\ 
\raisebox{0.1cm}{--\ 11}\end{array}$}
&
3
&
{\footnotesize $\begin{array}{l}\raisebox{-.05cm}{+11}\\ 
\raisebox{0.1cm}{--\ 10}\end{array}$}
&
8
& 7 & 4 & 8 & 4 & -- & -- & --
\\ \hline

784 &
{\footnotesize $\begin{array}{l}\raisebox{-.05cm}{+20}\\ 
\raisebox{0.1cm}{--\ 13}\end{array}$}
&
{\footnotesize $\begin{array}{l}\raisebox{-.05cm}{+13}\\ 
\raisebox{0.1cm}{--\ 11}\end{array}$}
&
3
&
{\footnotesize $\begin{array}{l}\raisebox{-.05cm}{+12}\\ 
\raisebox{0.1cm}{--\ 10}\end{array}$}
&
{\footnotesize $\begin{array}{l}\raisebox{-.05cm}{+ 9}\\ 
\raisebox{0.1cm}{-- \ 8}\end{array}$}
& 7 & 4 & 8 & 4 & -- & -- & --
\\ \hline

865 &
{\footnotesize $\begin{array}{l}\raisebox{-.05cm}{+22}\\ 
\raisebox{0.1cm}{--\ 14}\end{array}$}
&
{\footnotesize $\begin{array}{l}\raisebox{-.05cm}{+14}\\ 
\raisebox{0.1cm}{--\ 12}\end{array}$}
&
3
&
{\footnotesize $\begin{array}{l}\raisebox{-.05cm}{+13}\\ 
\raisebox{0.1cm}{--\ 11}\end{array}$}
&
{\footnotesize $\begin{array}{l}\raisebox{-.05cm}{+ 9}\\ 
\raisebox{0.1cm}{-- \ 8}\end{array}$}
& 8 & 4 & 9 & 4 & -- & -- & --
\\ \hline

968 &
{\footnotesize $\begin{array}{l}\raisebox{-0.05cm}{+24}\\ 
\raisebox{0.1cm}{--\ 15}\end{array}$}
&
{\footnotesize $\begin{array}{l}\raisebox{-0.05cm}{+15}\\ 
\raisebox{0.1cm}{--\ 13}\end{array}$}
&
3
&
{\footnotesize $\begin{array}{l}\raisebox{-0.05cm}{+14}\\ 
\raisebox{0.1cm}{--\ 12}\end{array}$}
&
{\footnotesize $\begin{array}{l}\raisebox{-0.05cm}{+ 9}\\ 
\raisebox{0.1cm}{-- \ 8}\end{array}$}
& 8 & 4 & 9 & 4 & -- & -- & --
\\ \hline
\end{tabular}
\end{center}
\end{table}

\clearpage

Table II: For each bin we list the average dijet mass, the 
differential cross section, and the statistical and total systematic 
uncertainty on the cross section.

\begin{table}[h]
\begin{center}
\begin{tabular}{|c|c|c|c|c|}\hline
bin edge & average $M$ & $d\sigma/dM$ & statistical  & systematic \\ 
(GeV/c$^2$) & (GeV/c$^2$)      & (pb/GeV/c$^2$)    & uncertainty  &  uncertainty  \\ \hline
180 & 188 & 6.07$\times 10^{2}$ &  3.2\,\% & $^{+20}_{-17}$\,\% \\
198 & 207 & 3.42$\times 10^{2}$ &  4.1\,\% & $^{+19}_{-17}$\,\% \\
217 & 228 & 1.81$\times 10^{2}$ &  1.0\,\% & $^{+19}_{-16}$\,\% \\
241 & 252 & 9.81$\times 10^{1}$ &  1.4\,\% & $^{+19}_{-16}$\,\% \\
265 & 277 & 4.98$\times 10^{1}$ &  1.8\,\% & $^{+19}_{-17}$\,\% \\
292 & 305 & 2.78$\times 10^{1}$ &  1.1\,\% & $^{+19}_{-17}$\,\% \\
321 & 335 & 1.43$\times 10^{1}$ &  1.4\,\% & $^{+20}_{-17}$\,\% \\
353 & 368 & 7.41$\times 10^{0}$ &  1.9\,\% & $^{+20}_{-18}$\,\% \\
388 & 405 & 3.83$\times 10^{0}$ &  0.9\,\% & $^{+21}_{-18}$\,\% \\
427 & 446 & 1.89$\times 10^{0}$ &  1.2\,\% & $^{+21}_{-19}$\,\% \\
470 & 491 & 9.07$\times 10^{-1}$ &  1.7\,\% & $^{+22}_{-19}$\,\% \\
517 & 539 & 4.50$\times 10^{-1}$ &  2.3\,\% & $^{+23}_{-20}$\,\% \\
568 & 592 & 1.90$\times 10^{-1}$ &  3.3\,\% & $^{+25}_{-21}$\,\% \\
625 & 652 & 7.42$\times 10^{-2}$ &  5.1\,\% & $^{+26}_{-22}$\,\% \\
688 & 716 & 2.92$\times 10^{-2}$ &  7.7\,\% & $^{+28}_{-23}$\,\% \\
756 & 784 & 1.18$\times 10^{-2}$ & 11\,\%   & $^{+30}_{-25}$\,\% \\
832 & 865 & 3.57$\times 10^{-3}$ & 20\,\%   & $^{+32}_{-26}$\,\% \\
915 & 968 & 9.03$\times 10^{-4}$ & 33\,\%   & $^{+34}_{-28}$\,\%\\ \hline
\end{tabular}
\end{center}
\end{table}

\clearpage

Table III: $\chi^2$ and corresponding probability for theoretical predictions 
for the dijet mass spectrum with various choices of parton distribution 
functions and renormalization scales $\mu=D E_T^{max}$. The row labeled
Fit is the parameterization used in the unsmearing (see text).

\begin{table}[h]
\renewcommand{\baselinestretch}{1.3}
\begin{center}
\begin{tabular}{|c|c|c|l|}\hline
PDF    & D  & $\chi^2$ & Probability \\ \hline
CTEQ4M  & 0.25 & 66.0 & $2.2 \times 10^{-7}$ \\   
        & 0.5 & 48.9 &  $1.1 \times 10^{-4}$ \\  
        & 1.0 & 48.1 &  $1.5 \times 10^{-4}$ \\  
        & 2.0 & 52.5 &  $1.7 \times 10^{-5}$ \\ 
CTEQ4HJ & 0.5 & 29.8 &  $4.0 \times 10^{-2}$ \\   
        & 1.0 & 26.1 &  $9.8 \times 10^{-2}$ \\  
CTEQ3M  & 0.5 & 45.7 &  $3.3 \times 10^{-4}$ \\   
        & 1.0 & 55.2 &  $1.2 \times 10^{-5}$ \\    
MRST    & 0.5 & 38.7 &  $3.2 \times 10^{-3}$ \\    
        & 1.0 & 33.5 &  $1.5 \times 10^{-2}$  \\   
MRST(g$\uparrow$) 
        & 0.5 & 36.1 &  $6.9 \times 10^{-3}$ \\    
MRST(g$\downarrow$)
        & 0.5 & 38.3 &  $3.5 \times 10^{-3}$ \\    
Fit     & --  & 20.5 &  $1.2 \times 10^{-1}$ \\     
\hline
\end{tabular}
\end{center}
\end{table}

\clearpage

\clearpage

\end{document}